\documentclass[aps,twocolumn,groupedaddress,floatfix,superscriptaddress]{revtex4}
\usepackage{graphicx}
\usepackage{amsmath}
\usepackage{amstext}
\usepackage{amssymb}
\usepackage{amsfonts}   
\usepackage{amsbsy}

\newcommand{\bpm}{\begin{pmatrix}}
\newcommand{\epm}{\end{pmatrix}}
\newcommand{\bmm}{\begin{matrix}}
\newcommand{\emm}{\end{matrix}}

\begin{document}
\bibliographystyle{apsrev}

\title{Gapless layered three-dimensional fractional quantum Hall states}

\author{Michael Levin}
\affiliation{Department of Physics, University of California, Santa 
Barbara, California 93106}
\author{Matthew P. A. Fisher}
\affiliation{Microsoft Research, Station Q, University of California, 
Santa Barbara, California 93106}

\date{\today}

\begin{abstract}
Using the parton construction, we build a three-dimensional (3D) multilayer fractional 
quantum Hall state with average filling $\nu = 1/3$ per layer that is 
qualitatively distinct from a stacking of weakly coupled Laughlin states.
The state supports gapped charge $e/3$ fermionic quasiparticles that 
can propagate both within and between the layers, in contrast
to the quasiparticles in a multilayer Laughlin state which are confined within each layer.
Moreover, the state has gapless neutral collective modes,
a manifestation of an emergent ``photon", which is minimally coupled to the 
fermionic quasiparticles. The surface sheath of the multilayer state resembles
a chiral analog of the Halperin-Lee-Read state, which is protected
against gap forming instabilities by the topological character of the bulk 3D 
phase. We propose that this state might be present in 
multilayer systems in the ``intermediate tunneling regime", where the 
interlayer tunneling strength is on the same order as the Coulomb energy 
scale. We also find that the parton construction leads to a candidate state for a 
bilayer $\nu = 1/3$ system in the intermediate tunneling regime. The candidate
state is distinct from both a bilayer of $\nu=1/3$ Laughlin states and the 
single layer $\nu = 2/3$ state, but is nonetheless a fully gapped fractional quantum 
Hall state with charge $e/3$ anyonic quasiparticles.
\end{abstract}

\maketitle

\section{Introduction}
While the fractional quantum Hall effect is firmly rooted in two dimensions 
(2D), \emph{anisotropic} 3D electron systems - such as multilayer 
systems in a perpendicular magnetic field - can exhibit fractional quantum 
Hall (FQH) states, at least in principle. The simplest example of such a 3D multilayer state is a 
stacking of $N$ decoupled Laughlin states. \cite{L8395, BF9682} At the next 
level of complexity, one can construct states with interlayer correlations, such 
as the $(3,3,1)$ bilayer state. \cite{H8375} More generally, multicomponent 
Chern-Simons theory allows one to construct a myriad of $N$-layer 
analogues of the $(3,3,1)$ state. \cite{NPS0008} 

These states are quite general but they suffer from a limitation: they all have a fixed
number of electrons in each layer. This restriction could be problematic for describing
certain multilayer systems, especially those with appreciable interlayer tunneling.
Therefore, alternative constructions of 3D multilayer FQH states are desirable theoretically.

On the experimental side, a number of experiments on 3D semiconductor multilayers 
have explored the behavior of 
stacked integer quantum Hall states \cite{WDD0412,DGM0305,DTM9865}, 
including the novel vertical transport due to the conducting surface sheath. 
\cite{BF9682,BFZ9701, CBF9714, CD9596, CS9999, TCC0507, TCC0504,BC0031} 
More recently, experiments on 
bismuth crystals in high magnetic fields have revealed intriguing anomalies 
in the ultra quantum limit - the limit where the magnetic field is 
sufficiently large that only the lowest Landau level is (partially) 
occupied. \cite{BBK0729} It has been suggested that a novel 3D fractional 
quantum Hall type state might be present. While these are not layered 
materials, strong electron correlations could drive a transition wherein the 
electrons spontaneously form a weak layered structure, as suggested in recent 
work. \cite{AB0861,BBA0910} Bulk graphite also reveals 
transport anomalies in the ultra quantum limit \cite{ID8582, OTT9286, YS9893}, 
which have been attributed 
to a charge density wave transition in this layered material. The observed 
quantum Hall effect in graphene \cite{NGM0597,ZTS0501,NMM0677, ZJS0606}, and 
future prospects for graphene multilayers, provides further impetus to 
explore 3D layered quantum Hall phenomena.

Motivated by these experiments, and the lack of previous theoretical 
exploration, we revisit the behavior of multilayer systems 
in the fractional quantum Hall regime. Generally, we are interested in 
addressing the following class of questions: what is the fate of a weakly 
coupled stacking of 2D fractional quantum Hall states when the interlayer 
electron tunneling becomes strong enough to close the quantum Hall gap? In 
particular, are new fractional quantum Hall type states possible when the 
Coulomb interaction is comparable to the interlayer tunneling strength? 
We believe that this intermediate tunneling strength regime is both 
experimentally accessible and theoretically novel.

Answering these questions definitively for a specific microscopic model is 
quite challenging and likely requires extensive numerical calculation. 
Here, we are less ambitious. Our goal is simply to find 
\emph{candidate} ground states for the intermediate tunneling 
regime. This is already a non-trivial problem since, as we mentioned earlier,
most multilayer states - 
such as those obtained from Chern-Simons mean field theory - have
a fixed number of electrons in each layer and hence are unnatural unless 
the interlayer tunneling is weak.

In this paper, we construct a candidate ground state for the simplest 
possible multilayer system: spinless (or spin polarized) electrons with an 
average filling of $\nu = 1/3$ per layer. We speculate that the candidate 
state may be realized at intermediate tunneling strength. However, our 
arguments for the candidate state are 
indirect, as we do not make any detailed analysis of energetics.

We construct our candidate state using a slave-particle gauge theory 
approach known as the ``parton construction." \cite{J8979,BW9033} The basic 
idea of the parton construction is to write the electron creation operator as 
a product of several (in our case, $3$) fermionic parton creation 
operators. By choosing different mean field parton states, and 
including (gauge) fluctuations, one can construct different FQH states.
The advantage of the parton construction is that it naturally leads to
states with electron number fluctuations in each layer. Thus, parton FQH 
states may be particularly natural in the intermediate tunneling regime.
In addition, the particular state we construct has the interesting property 
that one can tune from it to a decoupled multilayer Laughlin state by 
changing a single coupling constant in the parton gauge theory. 
Given that the multilayer Laughlin state is likely realized at weak 
interlayer tunneling, this is an additional reason we consider our state 
to be a candidate ground state for intermediate interlayer tunneling.

We analyze the candidate state for both a finite number of layers $N$, and 
for the 3D limit, $N \rightarrow \infty$. For a finite number of 
layers $N$, we find that the state is a fully gapped FQH
state (when $N=1$, it is simply the Laughlin state). The quasiparticle 
excitations are anyonic and carry charge $e/3$. We find that the quasiparticle 
excitations are described by a $K$-matrix  \cite{W9505} 
of dimension $(3N-2) \times(3N-2)$, along with a charge vector
of length $(3N-2)$. 

In the 3D limit, the candidate state exhibits more unusual physics. 
It supports two types of excitations: gapped charge $e/3$ fermionic 
quasiparticle excitations and gapless, electrically neutral collective 
modes. The $e/3$ fermionic quasiparticle excitations (which are 
essentially the ``deconfined" partons) are truly 3D quasiparticles and can 
move freely between layers. This should be contrasted with the $e/3$ 
excitations in the multilayer Laughlin state which are confined to 
individual layers. (In this sense, our candidate state is ``more 3D" then 
the multilayer Laughlin state). As for the gapless electrically neutral 
collective modes, these are the emergent $U(1)$ gauge bosons or ``photons" 
which originate from fluctuations about the mean field parton state. 
Unlike Maxwell photons, these excitations have only one polarization state 
and have an anisotropic dispersion of the form $\omega_{\bf k}^2 \sim {\bf 
k}_\perp^2 + k_z^4$ (for layers oriented in the $xy$ plane). The $e/3$ 
fermionic quasiparticles are minimally coupled to these ``photon" modes and 
thus have long range interactions.

The edge physics of the $N$-layer and 3D systems is also interesting.
In the case of a finite number of layers $N$, the edge theory is a chiral 
boson conformal field theory with $3N-2$ chiral modes. The edge 
Lagrangian can be read off from the bulk $K$-matrix using the standard 
formalism. \cite{W9505} The edge (or surface) physics in the 3D limit is more complex. 
For a 3D system with layers in the $xy$ plane and a boundary in the $xz$ 
plane, the three flavors of dispersing edge modes form a ``sheath"  of chiral 
fermions. At the mean field level these fermions are non-interacting, but 
with fluctuations included are minimally coupled to the gapless bulk 
``photons". The surface sheath, then, resembles a chiral analog of the 
Halperin-Lee-Read state, \cite{HLR9312} which is protected against gap-forming 
instabilities by the topological character of the bulk phase.

This paper is organized as follows. In section \ref{phasediagsect} we 
speculate about phase diagrams of multilayer $\nu = 1/3$ systems and we
describe the intermediate tunneling regime in more detail. In section 
\ref{partconstrsect} we construct our candidate state. In sections 
\ref{singlayersect} -\ref{3Dsect}, we analyze the bulk physics of the 
candidate state for single layer, bilayer, $N$-layer, and 3D systems. 
In section \ref{crossover} we investigate the crossover between
2D and 3D physics in systems with a large but finite number
of layers. In section \ref{declayersect} we describe the relationship between the 
candidate state and the multilayer Laughlin state. Finally, in section 
\ref{edgesect} we analyze the edge physics of the candidate state for 
$N$-layer and 3D systems.

\section{Model and possible phase diagrams}
\label{phasediagsect}
In this section, we discuss the physics of $N$-layer $\nu = 1/3$ FQH 
systems in more detail. We speculate about possible phase diagrams 
and we explain what the intermediate tunneling regime is, and why it is 
interesting.

Consider a geometry where the layers are oriented in the $xy$ plane,
and neighboring layers are spaced a distance $a$ in the $z$ 
direction. There are four energy scales in the  
problem - two intralayer and two interlayer scales. The intralayer energy 
scales are the cyclotron energy $\hbar \omega_c = \hbar e B/m$, and the 
characteristic intralayer Coulomb energy scale $E_c = e^2 /l_B$. The 
interlayer scales are the interlayer Coulomb energy $e^2/a$ 
and the interlayer tunneling strength $t_z$. In the following discussion, 
we will focus on the regime where (1) $\hbar \omega_c$ is much larger 
than any of the other energy scales, and (2) $a$ is comparable 
to, but larger than $l_B$ so that the interlayer Coulomb energy scale
is smaller than, but on the same order as the intralayer energy $E_c$.
In this regime, there is only one dimensionless parameter in the problem: 
the ratio $g \equiv t_z/E_c$. 

Let us think about the phase diagram as we vary the dimensionless ratio 
$g = t_z/E_c$. To begin, suppose $N=2$. When $g=0$, there is no interlayer 
tunneling and we expect that the ground state is given by two decoupled Laughlin $\nu=1/3$ states, 
with perhaps small quantitative modifications due to the interlayer 
Coulomb interaction. Since the Laughlin state is gapped, it will be stable
to small interlayer hopping, i.e. $g \ll 1$. In the opposite limit, with 
very large interlayer tunneling $g \gg 1$, all of the electrons will be in 
the ``bonding band" - the band consisting of symmetric combinations of
Landau orbitals in the two layers. The system is thus an effective
single layer system at filling $\nu=2/3$. The weak Coulomb 
interactions will presumably lead to an abelian $\nu=2/3$ state
with gap of order $E_c$.  

Now, consider the regime $g \sim 1$. Starting from the decoupled limit, 
when $g$ is increased one expects that the quasiparticle gap in each layer 
will diminish and presumably be driven to zero at some critical value 
$g_1$. On the other hand, when $g$ is brought down from large values, the 
gap of the ``single layer" $\nu=2/3$ state will diminish (due to Landau 
level mixing into the ``antibonding band") and be driven to zero at some 
value, $g_2$. Generally, there is no reason to expect that $g_1=g_2$, 
although it is possible that there is a direct first order transition 
between the $1/3 + 1/3$ decoupled phase and the $2/3$ ``single" layer 
state. If $g_2 > g_1$, there will be a third FQH phase
(or phases) for $g_1 < g < g_2$. This potential phase is the ``intermediate 
coupling" phase we consider in this paper.

One can imagine a similar scenario for the 3D limit, $N \rightarrow 
\infty$. Again, at weak interlayer tunneling, $g \ll 1$, the stack of 
Laughlin states is stable and can be readily analyzed. \cite{BF9682}
On the other hand, the nature of the strong tunneling phase with $g \gg 1$ 
is non-trivial. In the non-interacting limit, the lowest Landau level in 
each layer will form a band that disperses in the $z-$direction,
with energy $ -t_z \cos(k_z a)$. The noninteracting ground state
consists of completely filling all of the lowest Landau level band states 
with  $|k_z| < k_F \equiv \pi/3a$. This describes a gapless state. In a 
Wannier basis of orthonormalized lowest Landau level wavefunctions with 
guiding centers sitting on the sites of a regular 2D lattice (say 
triangular), one can view the system as an array of one-dimensional 
non-interacting electron systems. Gapless particle-hole excitations exist 
across the two Fermi points in each of the 1D  ``wires". This 
noninteracting state is likely to be unstable in the presence of 
arbitrarily weak Coulomb interactions, $E_c \ne 0$, due to the nested 
$2k_F$ backscattering interactions between nearby ``wires". The simplest 
scenario would be the development of a fully gapped $Q= 2k_F= 2\pi/(3a)$ 
charge density wave (CDW) state, which corresponds to a tripling of the unit cell along the 
$z-$axis. Naively one would expect the CDW gap to scale as 
$\Delta_{\text{CDW}} \sim t_z \exp(-\text{const.} \cdot g)$. \cite{LBF9769,LBF9894}
This CDW state can be loosely thought of as an effective system with 
$N/3$ layers, each one at filling $\nu_{eff}=1$.

As in the bilayer case, increasing $g$ from small values
will presumably close the Laughlin quasiparticle gap in each of the layers,
destroying the decoupled phase at some $g_1$. Similarly, when $g$ is 
decreased from very large values down to order one the CDW state will 
become disfavored due to the increasing intralayer Coulomb repulsion. One 
expects the CDW state to be destroyed for some $g < g_2$.  As for $N=2$, 
it is possible that there is a third phase for $g_1 < g < g_2$ - an 
``intermediate tunneling phase."

The possibility of such an intermediate tunneling phase for either the 
finite $N$ case or the 3D limit $N \rightarrow \infty$, is the starting 
point for this paper. One reason we feel it is a particularly interesting
possibility is that the interlayer tunneling and the Coulomb interaction 
are both of paramount importance in such a putative phase. This poses 
a theoretical challenge since the obvious FQH states - such as those 
constructed from Chern-Simons mean field theory - have a fixed number of 
electrons in each layer and are therefore unnatural except for very weak 
interlayer tunneling. On the other hand, if one tries to construct a state 
by treating the $N$-layer system as an effective single layer system at 
filling $N/3$, the result is unnatural except for very strong interlayer 
tunneling. In this paper, we use a different approach - a slave particle 
construction with fermionic ``partons" - to build a candidate state that 
overcomes these difficulties.

\section{Parton construction}
\label{partconstrsect}
Our candidate state can be described most naturally using the parton 
construction. \cite{J8979,BW9033} Let us first describe the construction in 
the case of the single layer system; we will then generalize to multiple 
layers.

Our starting point is the single layer electron Hamiltonian. As it will be
convenient in what follows, we regularize this Hamiltonian, replacing the 
2D continuum by a square lattice. We take the flux through each plaquette 
in the lattice to be $2\pi/M$, the electron density to be $1/3M$, and we 
consider the limit $M \rightarrow \infty$. In this limit, the lattice 
model behaves like a $2D$ continuum with electrons at filling fraction 
$\nu = 1/3$.

The regularized electron Hamiltonian can be written as
\begin{equation}
H = -\sum_{xi} (t c_{x}^\dagger e^{i \tilde{A}_{x,i}} c_{x + \hat{x}_i} + 
h.c. ) + \text{interactions}
\end{equation}
where $\tilde{A}_{x,i}$ is a lattice gauge field with $\Delta_1 
\tilde{A}_{x,2} - \Delta_2 \tilde{A}_{x,1}  = 2\pi/M$, and $\tilde{A}$ is 
periodic with unit cell of size $M$. Here, $\Delta_i$ denotes a lattice 
derivative in the $\hat{x}_i$ direction: $\Delta_i f_x \equiv 
f_{x+\hat{x}_i}-f_x$.

In the parton construction, we think of the electron as a composite of $3$ 
fermionic partons $d^p$, $p=1,2,3$:
\begin{equation}
c = d^1 d^2 d^3
\end{equation}
We then substitute the expression for $c$ into this Hamiltonian and 
expand around a saddle point. The result is a non-interacting mean-field 
Hamiltonian for the partons. Many different saddle 
points can be stabilized depending on the details of the interactions. 
Different saddle points correspond to different FQH states. 

Here, we consider a particular saddle point. The saddle point we are 
interested in is associated with the mean field Hamiltonian 
\begin{equation}
H_{mf} = -\sum_{xip}(t_{p} (d^p_{x})^\dagger e^{i A_{x,i}} d^p_{x + 
\hat{x}_i} + h.c.) 
\end{equation}
where $A_{x,i}$ is a lattice gauge field with $\Delta_1 A_{x,2} - 
\Delta_2 A_{x,1}  = 2\pi/3M$, and $A_{x,i}$ is periodic with unit cell of 
size $3M$. Notice that the flux $2\pi/3M$ is exactly the right size so 
that the partons are at filling $\nu = 1$ (the partons, like the 
electrons, are at density $1/3M$). We have also assumed that the hopping 
amplitudes $t_p$ are different for the three species of partons.  

What is the physics of this parton state? At the mean field level, the 
parton state is a gapped state with 
fermionic excitations. However, this mean field result is not quite 
correct, since we have not taken into account the effect of fluctuations 
about the saddle point. These fluctuations are described by fluctuations 
in the hopping amplitudes $t_p$ of the form $t_{p} \rightarrow t_p 
e^{i\theta_p}$ with $\theta_1 + \theta_2 + \theta_3 = 0$. We can 
parameterize them in terms of two $U(1)$ gauge fields $A^q$, $q=1,2$ by 
setting $\theta_p = Q_{pq} A^q$ where $q=1,2$, and $Q_{p1} = (1,0,-1)$, 
$Q_{p2} = (0,1,-1)$. The effect of fluctuations is thus to couple the 
partons to two $U(1)$ gauge fields $A^q$. (Note that the structure of the 
gauge fluctuations is closely tied to the symmetries of the saddle point. 
For example, at the symmetric saddle point $t_1 = t_2 = t_3$ the 
fluctuations are described by $SU(3)$ gauge fluctuations $t_p \rightarrow 
t_{p'} U^{p'}_p$ rather than the $U(1) \times U(1)$ fluctuations present 
here \cite{W9102}). Including these fluctuations, our Hamiltonian is given 
by
\begin{equation}
H = H_t + H_A
\label{singlayer}
\end{equation}
where $H_t$ describes the parton hopping and $H_A$ describes the gauge 
field dynamics:
\begin{eqnarray}
H_t &=& - \sum_{x i p} (t_p (d^p_{x})^\dagger e^{iQ_{pq} A^{q}_{x, i}+
i A_{x, i}} d^p_{x + \hat{x}_i} + h. c.) \\
H_A &=&  \sum_{x i q} \frac{g}{2} (E^{q}_{x, i})^2 
-\sum_{x q} J \cos(\Delta_1 A^{q}_{x, 2} -\Delta_2 A^{q}_{x, 1}) 
\end{eqnarray}
The parton state we are interested in is described by the above 
Hamiltonian in the weak gauge fluctuation regime - e.g. $g \ll J, t_p$.

Generalizing this construction to the $N$ layer case is straightforward. 
In this case, the mean field parton Hamiltonian is given by
\begin{eqnarray}
H_{mf} &=& -\sum_{xzip}(t_{p\perp} (d^p_{xz})^\dagger e^{i A_{x,i}} 
d^p_{(x + \hat{x}_i)z} + h.c.) 
\nonumber \\
&-& \sum_{xzp} (t_{p3} (d^p_{xz})^\dagger d^p_{x(z+1)} + h.c.)
\end{eqnarray}
where $z$ is the layer index and $t_{p\perp}, t_{p3}$ are the intralayer 
and interlayer hopping amplitudes.

Including the $U(1) \times U(1)$ gauge fluctuations, we arrive at
\begin{equation}
H = \sum_{z=1}^N (H_{zt}  + H_{zA}) + \sum_{z=1}^{N-1} (H_{z(z+1)t} 
+H_{z(z+1)A}) 
\label{multlayer}
\end{equation}
where $H_{zt}, H_{zA}$ describe the intralayer hopping and gauge field terms,
\begin{align}
&H_{zt} = - \sum_{x i p} (t_{p\perp} (d^p_{x z})^\dagger e^{iQ_{pq} 
A^{q}_{x z,i}+i A_{x,i}} d^p_{(x + \hat{x}_i)z} + h. c.) \\
&H_{zA} =  \sum_{x i q} \frac{g_\perp}{2} (E^{q}_{x z,i})^2 - 
\sum_{x q} J_{\perp} \cos(\Delta_1 A^{q}_{x z,2} -\Delta_2 A^{q}_{x z,1})
\end{align}
and $H_{z(z+1)t}, H_{z(z+1)A}$ describe the interlayer hopping and gauge 
field terms,
\begin{align}
&H_{z(z+1) t} = - \sum_{x p}( t_{p3} (d^p_{x z})^\dagger e^{i Q_{p q} 
A^{q}_{x z,3}} d^p_{x (z+1)} + h.c.) \nonumber \\
&H_{z(z+1)A} = \sum_{x q} \frac{g_3}{2} (E^{q}_{xz,3})^2  \nonumber \\
&- \sum_{xiq} J_3 \cos(\Delta_i A^{q}_{xz,3}- A^{q}_{x z,i} + A^{q}_{x 
(z+1),i}) 
\label{interlaygauge}
\end{align}
Again, we assume that the gauge fluctuations are weak - $g_3 \ll J_3, 
t_{p3},t_{p\perp}$ and $g_\perp \ll J_\perp, t_{p3}, t_{p\perp}$. In the 
following sections, we analyze the physics of this state. 

\section{Single layer}
\label{singlayersect}
We begin with the simplest case: the single layer parton state. We 
rederive the well-known result that the single layer parton state is 
precisely the Laughlin state. \cite{J8979,BW9033} 

To understand the properties of the single layer parton state, we need to 
analyze the low energy physics of the Hamiltonian (\ref{singlayer}).
One way to do this is to introduce $U(1)$ gauge fields $a^p_\mu$ to describe 
the parton number currents:
\begin{equation}
j^{p \lambda} = \frac{1}{2\pi} \epsilon^{\lambda \mu \nu} \partial_\mu a^p_\nu
\end{equation}
The low energy effective theory for the parton hopping terms $H_t$ can 
then be written as
\begin{equation}
L = \frac{1}{4\pi} \sum_p \epsilon^{\lambda \mu \nu} a^p_\lambda 
\partial_\mu a^p_\nu + \text{minimal coupling to } A^q
\end{equation}
Including the minimal coupling to $A^q$ gives
\begin{equation}
L = \frac{1}{4\pi} \sum_p \epsilon^{\lambda \mu \nu} a^p_\lambda 
\partial_\mu a^p_\nu + \frac{1}{2\pi} \epsilon^{\lambda \mu \nu} Q_{pq} A^q_{\lambda} 
\partial_\mu a^p_{\nu}
\end{equation}
Adding the gauge field terms $H_A$, expanding the cosines to quadratic 
order, and taking the continuum limit, we arrive at the low energy 
effective theory
\begin{eqnarray}
L &=& \frac{1}{4\pi} \sum_p \epsilon^{\lambda \mu \nu} a^p_\lambda 
\partial_\mu a^p_\nu + 
\frac{1}{2\pi} \epsilon^{\lambda \mu \nu} Q_{pq} A^q_{\lambda} 
\partial_\mu a^p_{\nu}
\nonumber \\
&+& \sum_{iq} \frac{1}{2g} (\partial_0 A^{q}_{i} - \partial_i 
A^{q}_{0})^2 \nonumber \\
&-& \sum_{q} \frac{J l^2}{2} (\partial_1 A^{q}_{2} - \partial_2 A^{q}_{1})^2 
\end{eqnarray}
where $l$ is the lattice spacing.
The last two terms are irrelevant to the low energy physics since 
integrating out the $a^p_\mu$ field produces a Chern-Simons term for $A^q$ 
(which has one less derivative then the above Maxwell terms). Dropping 
these terms and integrating out $A$ leaves us with
\begin{equation}
L = \frac{1}{4\pi} \sum_p \epsilon^{\lambda \mu \nu} a^p_\lambda 
\partial_\mu a^p_\nu
\end{equation}
together with the constraints $\partial_\mu a^1_{\nu} = \partial_\mu 
a^2_{\nu} = \partial_\mu a^3_{\nu}$. Letting $a_\nu = a^1_{\nu} = a^2_{\nu} 
= a^3_{\nu}$ we get
\begin{equation}
L = \frac{1}{4\pi} \epsilon^{\lambda \mu \nu} 3 a_{\lambda} \partial_\mu
a_{\nu}
\end{equation}
If we include the coupling to the physical electromagnetic gauge field 
$A_{EM}$ - assigning electric charges $e_1,e_2,e_3$ to the partons with $e_1 
+ e_2 + e_3 = e$ - we find
\begin{equation}
L = \frac{1}{4\pi} \epsilon^{\lambda \mu \nu} 3 a_{\lambda} \partial_\mu
a_{\nu} + \frac{e}{2\pi} \epsilon^{\lambda \mu \nu} A_{EM,\lambda} 
\partial_\mu a_\nu
\end{equation}
(irrespective of the values of $e_1,e_2,e_3$). This is the low energy 
effective theory for the Laughlin state. We conclude that the single 
layer parton state is in the same universality class (e.g. quantum phase) 
as the Laughlin state.

\section{Bilayer}
\label{bilayersect}
In this section we analyze the parton state in the next simplest case: a 
bilayer. In this case, the parton Hamiltonian (\ref{multlayer}) reduces to
\begin{equation}
H = \sum_{z=1}^2 (H_{zt} + H_{zA}) + (H_{12t} + H_{12A}) 
\label{bilayer}
\end{equation}
where $H_{zt}, H_{zA}$ describe the intralayer hopping and gauge field terms,
\begin{align}
&H_{zt} = - \sum_{x i p} (t_{p\perp} (d^p_{x z})^\dagger e^{iQ_{pq} 
A^{q}_{x z, i}+
i A_{x, i}} d^p_{(x + \hat{x}_i) z} + h. c.) \\
&H_{zA} =  \sum_{x i q} \frac{g_\perp}{2} (E^{q}_{x z, i})^2 - 
\sum_{x q} J_{\perp} \cos(\Delta_1 A^{q}_{ x z, 2} -\Delta_2 A^{q}_{x z, 1})
\end{align}
and $H_{12t}, H_{12A}$ describe the interlayer hopping and gauge field terms,
\begin{eqnarray}
H_{12 t} &=& - \sum_{x p}( t_{p3} (d^p_{x 1})^\dagger e^{i Q_{p q} A^{q}_{x 
1, 3}} d^p_{x 2} + h.c.) \\
H_{12A} &=& \sum_{x q} \frac{g_3}{2} (E^{q}_{x1,3})^2 \nonumber \\
&-& \sum_{xiq} J_3 \cos(\Delta_i A^{q}_{x1,3} - A^{q}_{x 1, i} + A^{q}_{x 2, i}) 
\label{bilaygauge}
\end{eqnarray}
As before, we can derive the low energy physics of this Hamiltonian by 
introducing $U(1)$ gauge fields 
$a^p_{z\mu}$ to describe the parton number currents in each layer:
\begin{equation}
j^{p \lambda}_z = \frac{1}{2\pi} \epsilon^{\lambda \mu \nu} \partial_\mu 
a^p_{z\nu}
\end{equation}
Expanding the cosines in the gauge field terms to quadratic order, and 
putting everything together, we arrive at the effective theory
\begin{eqnarray}
L &=&  \frac{1}{4\pi} \sum_{z p} \epsilon^{\lambda \mu \nu} a^p_{z, \lambda} 
\partial_\mu a^p_{z, \nu} +  
\frac{1}{2\pi} \sum_{z} \epsilon^{\lambda \mu \nu} Q_{p q} A^q_{z, \lambda} 
\partial_\mu a^p_{z, \nu} \nonumber \\
&+& \frac{1}{2 g_3 a^2} (\partial_0 A^{q}_{1,3} - A^{q}_{1,0} + A^{q}_{2,0} 
)^2 \nonumber \\
&-& \sum_{iq} \frac{J_3}{2}(\partial_i A^{q}_{1,3} - A^{q}_{1,i} + 
A^{q}_{2,i})^2 
\label{bilayth}
\end{eqnarray}
where $a$ is the layer spacing. (As in the single layer case, we have dropped
the intralayer Maxwell terms as they are irrelevant to the low energy physics). 
To proceed further, we choose the gauge 
$A^q_{1,3} = 0$, and define new fields $A^{q \pm} = A^{q}_1 \pm A^{q}_2$. 
Expressing the Lagrangian in terms of these fields gives
\begin{eqnarray}
L &=&  \frac{1}{4\pi} \sum_{z p} \epsilon^{\lambda \mu \nu} a^p_{z, \lambda} 
\partial_\mu a^p_{z, \nu} \nonumber \\
&+& \frac{1}{4\pi} \sum_{\pm} \epsilon^{\lambda \mu \nu} Q_{p q} 
(A^{q\pm}_{\lambda} 
\partial_\mu (a^p_{1, \nu} \pm a^p_{2,\nu})) \nonumber \\
&+& \frac{1}{2 g_3 a^2} (A^{q-}_{0} )^2 - \sum_{iq} \frac{J_3}{2}(A^{q-}_{i})^2 
\end{eqnarray}
As in the single layer case, the final step is to integrate out the gauge 
fields $A^{q\pm}$. Integrating out $A^{q+}$ generates the constraints 
$\sum_z \partial_\mu a^1_{z,\nu} =  \sum_z \partial_\mu a^2_{z,\nu} = 
\sum_z \partial_\mu a^3_{z,\nu}$; integrating out $A^{q-}$ generates a 
Maxwell term for $a^p$ which is irrelevant to the low energy physics due 
to the presence of the Chern-Simons term.

We thus arrive at the Lagrangian
\begin{equation}
L =  \frac{1}{4\pi} \sum_{z p} \epsilon^{\lambda \mu \nu} a^p_{z, \lambda} 
\partial_\mu a^p_{z, \nu}
\end{equation}
together with the constraints $\sum_z \partial_\mu a^1_{z,\nu} =  \sum_z 
\partial_\mu a^2_{z,\nu} = \sum_z \partial_\mu a^3_{z,\nu}$.
There are four independent gauge fields left which we can parameterize by 
$a^1 = a^1_1, a^2 = a^2_1, a^3 = a^3_1, a^4 = \sum_z a^1_z = \sum_z a^2_z = 
\sum_z a^3_z$.
In terms of these variables, we have
\begin{equation}
L = \frac{1}{4\pi} \epsilon^{\lambda \mu \nu} K_{IJ} a^I_\lambda 
\partial_\mu a^J_\nu
\end{equation}
where
\begin{equation}
K = \bpm 2 & 0 & 0 & -1 \\
	        0 & 2 & 0 & -1 \\
	        0 & 0 & 2 & -1 \\
	        -1&-1&-1 &  3 \epm
\end{equation}	        
Including the coupling to the physical electromagnetic gauge field 
$A_{EM}$, assigning charges $e_1, e_2, e_3$ to the partons with $e_1 + e_2 
+ e_3 = e$, we find
\begin{equation}
L = \frac{1}{4\pi} \epsilon^{\lambda \mu \nu} K_{IJ} a^I_\lambda 
\partial_\mu a^J_\nu + \frac{e}{2\pi} \epsilon^{\lambda\mu\nu} t_I 
A_{EM,\lambda} \partial_\mu a^I_\nu
\end{equation}
where $t^T = (0,0,0,1)$ (irrespective of the values of $e_1, e_2, e_3$).

The parton state is completely specified by the above $K$-matrix and 
charge vector $t$ (or more accurately, the \emph{universal} properties of 
this quantum state are completely specified).
We now analyze the basic properties of this state.

We begin with the quasiparticle statistics. According to the $K$-matrix 
formalism, the quasiparticle excitations can be labeled by integer 
vectors $l$. The exchange statistics of a quasiparticle $l$ is given by 
$\theta_{ex} = \pi(l^T K^{-1} l)$. The mutual statistics of two 
quasiparticles $l$,$l'$ - e.g. the phase associated with braiding one 
particle around another - is given by $\theta_{mut} = 2\pi(l^T K^{-1} l')$. 

In principle, these formulas completely specify the quasiparticle 
statistics of the parton state. However, it is convenient to describe the 
statistics of the parton state in a more concise way. A general way to do 
this is to find a subset of quasiparticles with the property that one can 
generate all topologically distinct quasiparticles by taking composites 
of these basic quasiparticles. One can then describe the complete 
quasiparticle statistics by specifying the statistics of this generating 
subset of quasiparticles. For the above state, the three parton 
excitations corresponding to $l_1 = (1,0,0,0), l_2 = (0,1,0,0), l_3 = 
(0,0,1,0)$ generate all the others. (One way to see this is to note that 
the excitation $(0,0,0,1)$ is topologically identical to $(2,0,0,0)$). 
Simple algebra shows that the three parton excitations have exchange statistics
\begin{equation}
\theta_p = \pi (l_p^T K^{-1} l_p) = \frac{2\pi}{3}
\end{equation}
and mutual statistics
\begin{equation}
\theta_{pp'} = 2\pi (l_p^T K^{-1} l_{p'}) = \frac{(1+3\delta_{pp'})\pi}{3}
\end{equation}
This gives a complete description of the quasiparticle statistics of the 
bilayer parton state. The electric charges of the quasiparticles are also 
easy to obtain. Again, it suffices to specify the parton charges, which are given by
\begin{equation}
q_p = e \cdot (t^T K^{-1} l_p) = \frac{e}{3}
\end{equation}

Now that we have computed these properties, we can see that the bilayer 
parton state is distinct from a bilayer of decoupled $\nu = 1/3$ Laughlin 
states as well as the conventional $\nu = 2/3$ state. Indeed, one can 
distinguish the states by noting that the $e/3$ excitation in the parton 
state has a statistical angle $2\pi/3$, while the $e/3$ excitation in 
the other two states has an angle $\pi/3$.

One can also distinguish the states by their ground state degeneracy on a 
torus. This quantity is particularly easy to measure in numerical 
calculations. The ground state degeneracy for the bilayer parton state is 
just the determinant of $K$ which is $12$. On the other hand, the 
degeneracy of a bilayer of decoupled $\nu = 1/3$ Laughlin states is $9$, 
and the degeneracy of the $\nu =2/3$ state is $3$. 

A final way to distinguish the states is via their thermal Hall 
conductances. Recall that each chiral boson edge mode gives a contribution
of $\pm \frac{\pi^2 k_B^2}{3h} T$ to the thermal Hall conductance, with
the sign determined by the chirality of the mode. Thus, the thermal Hall
conductance can be computed by counting the number of positive and 
negative eigenvalues of $K$. In the case of the bilayer parton state,
there are four positive eigenvalues so the thermal Hall conductance is $4$ 
(in units of $\frac{\pi^2 k_B^2}{3h} T$). On the other hand, the thermal 
Hall conductance for the bilayer of Laughlin states is $2$ and the 
thermal Hall conductance for the $\nu = 2/3$ state is $0$.

\section{$N$-layer system}
\label{Nlayersect}
The bilayer results can be easily generalized to the $N$-layer case. For 
general $N$, one finds a $K$-matrix of dimension $(3N-2) \times 
(3N-2)$. The result is shown below for the case $N=3$:
\begin{equation}
K = \bpm 2 & 1 & 0 & 0 & 0 & 0 & -1 \\
         1 & 2 & 0 & 0 & 0 & 0 & -1 \\
         0 & 0 & 2 & 1 & 0 & 0 & -1 \\
         0 & 0 & 1 & 2 & 0 & 0 & -1 \\
	 0 & 0 & 0 & 0 & 2 & 1 & -1 \\
	 0 & 0 & 0 & 0 & 1 & 2 & -1 \\
	-1 &-1 &-1 &-1 &-1 &-1 & 3  \epm
\end{equation}
The corresponding charge vector is $t^T = (0,0,0,0,0,0,1)$. The
generalization to arbitrary $N$ is clear: along the diagonals there
are three $(N-1) \times (N-1)$ blocks of the form 
$1+\delta_{ij}$, while the last row and column is made up of $-1$'s
with a $3$ in the bottom right-hand corner.

As before, the $K$-matrix and charge vector determine all the universal 
properties of the FQH state, such as the quasiparticle statistics and 
charges. Also, just as before, one can summarize the quasiparticle charges 
and statistics more concisely by specifying the statistics and charges of the 
three parton species (which correspond to $l_1 = (1, 0,0,0,0,0, 0)$, $l_2 
= (0, 0, 1,0,0,0, 0)$, $l_3 = (0,0,0,0,1,0,0)$ in the $N=3$ case).

One finds that the parton excitations have exchange statistics
\begin{equation}
\theta_p = \frac{(3N-2)\pi}{3N}
\end{equation}
mutual statistics
\begin{equation}
\theta_{pp'} = \frac{(2+(6N-6)\delta_{pp'})\pi}{3N}
\end{equation}
and charge
\begin{equation}
q_p = \frac{e}{3}
\label{Ncharge}
\end{equation}
	        
\section{3D limit}
\label{3Dsect}
In this section, we analyze the parton construction in the $3D$ limit, $N 
\rightarrow \infty$. Recall that the parton Hamiltonian is given by
\begin{equation}
H = \sum_{z} (H_{zt} + H_{zA} + H_{z(z+1)t} + H_{z(z+1)A}) 
\end{equation}
where
\begin{equation}
H_{zt} = - \sum_{x i p} (t_{p\perp} (d^p_{x z})^\dagger e^{iQ_{pq} 
A^{q}_{x z,i}+i A_{x,i}} d^p_{(x + \hat{x}_i)z} + h. c.)
\end{equation}
and
\begin{equation}
H_{zA} =  \sum_{x i q} \frac{g_\perp}{2} (E^{q}_{x z,i})^2 - 
\sum_{x q} J_{\perp} \cos(\Delta_1 A^{q}_{x z,2} -\Delta_2 A^{q}_{x z,1})
\end{equation}
and
\begin{eqnarray}
H_{z(z+1) t} &=& - \sum_{x p}( t_{p3} (d^p_{x z})^\dagger e^{i Q_{p q} 
A^{q}_{x z,3}} d^p_{x (z+1)} + h.c.) \nonumber \\
H_{z(z+1)A} &=& \sum_{x q} \frac{g_3}{2} (E^{q}_{xz,3})^2 \\
&-& \sum_{xiq} J_3 \cos(\Delta_i A^{q}_{xz,3} -A^{q}_{x z,i} + A^{q}_{x 
(z+1),i}) \nonumber
\end{eqnarray}
As usual, we derive a low energy effective theory by introducing $U(1)$ 
gauge fields $a^p_{z\mu}$ to describe the parton number currents in each 
layer:
\begin{equation}
j^{p \lambda}_z = \frac{1}{2\pi} \epsilon^{\lambda \mu \nu} \partial_\mu 
a^p_{z\nu}
\end{equation}
Expanding the cosines in the gauge field terms to quadratic order, and 
putting everything together, we arrive at the effective Lagrangian
 \begin{equation}
L = L_{a} + L_{A} + L_{aA}
\end{equation}
where
\begin{equation}
L_a = \frac{1}{4\pi}\sum_{zp} \epsilon^{\lambda \mu \nu} a^p_{z,\lambda} 
\partial_\mu a^p_{z,\nu}
\end{equation}
and
\begin{eqnarray}
L_A &=& \sum_{ziq} \frac{1}{2g_\perp} (\partial_0 A^{q}_{z,i} - \partial_i 
A^{q}_{z,0})^2 \nonumber \\
&-& \sum_{zq} \frac{J_\perp l^2}{2} (\partial_1 A^{q}_{z,2} - 
\partial_2 A^{q}_{z,1})^2 \nonumber \\
&+& \sum_{zq} \frac{1}{2g_3 a^2} (\partial_0 
A^{q}_{z,3} - A^{q}_{z,0} + A^{q}_{z+1,0})^2 \nonumber \\
&-& \sum_{ziq} \frac{J_3}{2}(\partial_i A^{q}_{z,3} - A^{q}_{z,i} + 
A^{q}_{z+1,i})^2 
\end{eqnarray}
and
\begin{equation}
L_{aA} = \frac{1}{2\pi} \sum_z \epsilon^{\lambda \mu \nu} Q_{pq} 
A^q_{z,\lambda} \partial_\mu 
a^p_{z,\nu}
\end{equation}
To proceed further, we integrate out the gauge fields $a^p_z$ corresponding
to the parton excitations. Note that this is different from the approach we took
in the single layer and bilayer cases where we integrated out the gauge fields
$A^q$ instead. We could have used this approach in those cases as well. 
The advantage of this approach is that it leads 
to a simpler description of the bulk low energy physics: the low energy 
effective theory for the $N$ layer case is simply a $2 \times 2$ 
Chern-Simons theory coupled to fermionic partons (instead of a $(3N-2) 
\times (3N-2)$ Chern-Simons theory coupled to bosons). The disadvantage is 
that the edge physics cannot be easily
read off from the bulk effective theory. Here our primary interest is in 
the bulk physics - thus we choose to integrate out the fields $a^p_z$.

Integrating out the $a$ field (e.g. the partons) produces a Chern-Simons 
term for the $A$ field. We can then drop the $J_\perp$ and $g_\perp$ 
Maxwell terms, as they are irrelevant at long distances. The resulting
Lagrangian is given by
\begin{eqnarray}
L_{A,eff} &=& \frac{1}{4\pi}\sum_z \epsilon^{\lambda \mu \nu} K_{qq'} 
A^q_{z,\lambda} 
\partial_\mu A^{q'}_{z,\nu} \nonumber \\
&+& \sum_{zq} \frac{1}{2g_3 a^2} (\partial_0 A^{q}_{z,3} - A^{q}_{z,0} + 
A^{q}_{z+1,0})^2 \nonumber \\
&-& \sum_{ziq} \frac{J_3}{2}(\partial_i A^{q}_{z,3} - A^{q}_{z,i} + 
A^{q}_{z+1,i})^2 
\label{Nlaygaugeth}
\end{eqnarray}
where
\begin{equation}
K_{qq'} = \bpm -2 & -1 \\ -1 & -2 \epm
\end{equation}
The full low energy effective theory is described by fermionic partons 
minimally coupled to this gauge theory with gauge charges $Q_{pq}$:
\begin{equation}
L = L_{part} + L_{A,eff}
\end{equation}
where
\begin{eqnarray}
L_{part} &=& \sum_{pz} (d^p_z)^\dagger (i \partial_0 + Q_{pq} A^q_{z,0}) d^p_z \nonumber 
\\
&-&\sum_{ipz} \frac{1}{2m_{p\perp}} (d^p_z)^\dagger (\partial_i - iQ_{pq} A^q_{z,i})^2 
d^p_z \nonumber \\
&-& \sum_{pz} t_{p3} (d^p_z)^\dagger e^{iQ_{pq} A^q_{z,3}} d^p_{z+1}
+h.c.
\end{eqnarray}
(Here, $m_{p\perp} = 1/(2t_{p\perp} l^2)$).
We now analyze the physics of this low energy effective theory. We begin 
with the excitations in the bulk. There are two types of excitations: 
gapped parton excitations described by $d^p$ and gapless gauge boson 
excitations described by the above $U(1) \times U(1)$ gauge theory.

Let us try to understand the gapless gauge boson excitations in greater 
detail. We can derive the dispersion relation for these gapless modes by 
going to Fourier space. Going to Fourier space and taking $k$ small, we have
\begin{eqnarray}
L &=& \frac{1}{4\pi a} \epsilon^{\lambda \mu \nu} K_{qq'} A^q_{\lambda} ik_\mu 
A^{q'}_{\nu}
+ \sum_{q} \frac{1}{2g_3 a} (k_0 A^{q}_{3} - k_3 A^{q}_{0})^2 \nonumber \\
&-& \sum_{iq} \frac{J_3 a}{2}(k_i A^{q}_{3} - k_3 A^{q}_{i})^2 
\end{eqnarray}

Defining ${A}^{\pm} = \frac{1}{\sqrt{2}} (A^1 \pm A^2)$, $K$ is 
diagonalized and our Lagrangian becomes
\begin{eqnarray}
L &=& \frac{1}{4\pi a}\sum_q \epsilon^{\lambda \mu \nu} m_{q} A^q_{\lambda} 
ik_\mu A^{q}_{\nu} \nonumber \\
&+& \sum_{q} \frac{1}{2g_3 a} (k_0 A^{q}_{3} - k_3 A^{q}_{0})^2 \nonumber \\
&-& \sum_{iq} \frac{J_3 a}{2}(k_i A^{q}_{3} - k_3 A^{q}_{i})^2 
\end{eqnarray}
where $m_{\pm} = -3,-1$ are the two eigenvalues of $K$.
 
We can write this as
\begin{equation}
L = \sum_q (A^q)^\dagger M_q (A^q)
\end{equation}
where
\begin{equation}
M_q = \bpm \frac{k_3^2}{2g_3 a} & \frac{im_q k_2}{4\pi a} &  -\frac{im_q 
k_1}{4\pi a} & -\frac{k_0 k_3}{2 g_3 a} \\
		-\frac{im_q k_2}{4\pi a} & -\frac{J_3 a k_3^2}{2} & 
\frac{im_q k_0}{4\pi a} & \frac{J_3 a k_1 k_3}{2} \\
		\frac{im_q k_1}{4 \pi a} & -\frac{im_q k_0}{4\pi a} & 
-\frac{J_3 a k_3^2}{2} & \frac{J_3 a k_2 k_3}{2} \\
		-\frac{k_0 k_3}{2 g_3 a} & \frac{J_3 a k_1 k_3}{2} & 
\frac{J_3 a k_2 k_3}{2} & \frac{k_0^2}{2g_3 a} - \frac{J_3 a(k_1^2 + 
k_2^2)}{2} \epm
\end{equation}		 

Choosing the temporal gauge, $A_0 = 0$, we can reduce $M_q$ to the $3 
\times 3$ submatrix
\begin{equation}
M_q = \bpm  -\frac{J_3 a k_3^2}{2} & \frac{im_q k_0}{4\pi a} & \frac{J_3 
a k_1 k_3}{2} \\
		 -\frac{im_q k_0}{4\pi a} & -\frac{J_3 a k_3^2}{2} & 
\frac{J_3 a k_2 k_3}{2} \\
		 \frac{J_3 a k_1 k_3}{2} & \frac{J_3 a k_2 k_3}{2} & 
\frac{k_0^2}{2g_3 a} - \frac{J_3 a(k_1^2 + k_2^2)}{2} \epm
\end{equation}		 
Setting the determinant to $0$ we find one gapless mode for each $q = 
\pm$ with dispersion
\begin{equation}
\omega^2 = J_3 g_3 a^2 (k_1^2 + k_2^2) + \frac{4 \pi^2 J_3^2 a^4}{m_q^2} k_3^4
\label{disprel}
\end{equation}
In principle, these gapless modes should be visible in inelastic light scattering
measurements. Thus, such measurements could be used to distinguish the $3D$ parton
state from other candidate states - such as the decoupled Laughlin state. In
section \ref{edgesect} we describe another experimental signature of the
$3D$ parton state, involving surface excitations.

In addition to the gapless gauge excitations, the $3D$ state also 
contains gapped parton excitations. These excitations are fermions, and 
carry electric charge $q = e/3$ as in the $N$-layer case (\ref{Ncharge}). 
They are minimally coupled to the gapless gauge bosons and therefore have 
long range interactions. Note that these excitations are truly 3D 
quasiparticles - they can propagate freely both within layers and between 
layers. This should be contrasted with the charge $e/3$ particles in the 
decoupled Laughlin state, which are confined to individual layers. In this 
sense, the parton state is ``more 3D" than the decoupled Laughlin 
state.

\section{Crossover from 2D to 3D}
\label{crossover}
In the previous sections, we referred to the 3D limit as the limit $N 
\rightarrow \infty$. However, the 3D limit can also be accessed when there 
are a finite number of layers - provided that we probe the system at 
appropriate length and energy scales. In this section, we discuss this 
crossover from 2D to 3D physics.

Consider an $N$-layer system with $N \gg 1$. This system is described by
three species of partons minimally coupled to the the gauge theory 
(\ref{Nlaygaugeth}). According to the analysis following 
(\ref{Nlaygaugeth}), the low energy modes of this gauge theory satisfy the 
dispersion relation (\ref{disprel}). Since $N$ is finite, $k_3$ is quantized in 
multiples of $\pi/Na$. For each value of $k_3$, there is a corresponding 
2D mode. 

The mode with the smallest gap corresponds to $k_3 = \pi/Na$; the 
dispersion relation for this mode is
\begin{equation}
\omega^2 = J_3 g_3 a^2 (k_1^2 + k_2^2) + \frac{4 J_3^2 \pi^6}{m_q^2 N^4}
\end{equation}
We see that this mode has a gap $\Delta \sim  J_3/N^2$ and a correlation 
length $\xi \sim N^2 a \sqrt{g_3/J_3}$.

The gap $\Delta$ and correlation length $\xi$ are the important energy and 
length scales in the 2D/3D crossover. If one probes the system at energies 
less than $\Delta$ or lengths larger then $\xi$ (parallel to the layers), 
all the modes with $k_3 \neq 0$ will freeze out and the system will
behave like a gapped 2D system. The physics is then described by 
the gapped FQH state in section \ref{Nlayersect}. On the other hand, if one 
probes the system at energies greater than $\Delta$ or lengths smaller then 
$\xi$, the system will behave like a 3D system. In this case, the physics is 
described by partons coupled to the gapless gauge theory (\ref{Nlaygaugeth}).

\begin{figure}[tb]
\centerline{
\includegraphics[width=3.0in]{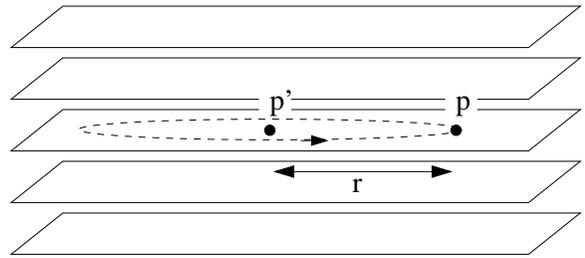}
}
\caption{
If $r \gg \xi$, the Berry phase associated with braiding a 
parton $p$ around another parton $p'$ is given by the 2D formula 
$\theta_{pp'} = 2\pi/3N$. If $r \ll \xi$, it depends on the 
details of the path and can be calculated from the 3D gapless gauge 
theory (\ref{Nlaygaugeth}).
}
\label{crossoverfig}
\end{figure}

One example of this crossover is the following thought 
experiment. Imagine adiabatically braiding one charge $e/3$ parton 
excitation $p$ around another $e/3$ parton excitation $p'$ - say of a 
different species - using a braiding path parallel to the layers (see 
Fig. \ref{crossoverfig}). First, 
consider the case where the separation $r$ between the partons is kept 
larger than $\xi$. In this case, the modes (\ref{disprel}) will be 
effectively frozen 
at this distance. While the presence of parton $p'$ will change the gauge 
flux seen by parton $p$, the gauge flux will be localized to within a 
distance $\xi$ of $p'$, and will be exponentially suppressed near the 
braiding path. Thus, the only interaction between the two partons will be 
a statistical interaction: the presence of parton $p'$ will change the 
total gauge flux enclosed by the braiding path of parton $p$. The 
Berry phase associated with braiding one parton around the other is then 
given by the 2D mutual statistics formula: $\theta_{pp'} = 2\pi/3N$.

Now, consider the case where the separation $r$ between the partons is 
kept 
much smaller than $\xi$. In this case, the modes (\ref{disprel}) will not be
frozen out and the partons will experience long range interactions. The 
phase associated with braiding one parton around the other will depend on 
the details of the path, and can be calculated using the 3D gapless gauge 
theory (\ref{Nlaygaugeth}). 

As for the crossover between the two regimes, we expect that the Berry 
phase in the 3D regime scales with the separation between the partons 
according to some power law. When the separation is of order $\xi$, we 
expect that the phase is of order $1/N$ so that it agrees with the phase 
in the 2D regime.

\section{Relationship to decoupled $\nu = 1/3$ layers}
\label{declayersect}
An interesting feature of the parton construction is that it can describe 
the decoupled $\nu = 1/3$ layered state within the same framework as the
3D parton state. One can tune from one state to the other by changing a single 
coupling constant in the parton gauge theory.

To see this, let us go back to the original parton Hamiltonian
(\ref{multlayer}). So far we have analyzed the physics of this  
Hamiltonian in the limit of weak gauge fluctuations: $g_3 \ll J_3,
t_{p3},t_{p\perp}$ and $g_\perp \ll J_\perp, t_{p3}, t_{p\perp}$. We found
that in this regime, the low energy physics was described by the 3D parton
state.

However, by increasing $g_3$ one can also access a regime where (interlayer) 
gauge fluctuations are strong. More specifically, suppose that  $g_3 \gg J_3, t_{p3},t_{p\perp}$. In this 
case, the compactness of the $U(1)$ gauge field becomes important: since 
the lattice electric field is integer valued and $g_3$ is large, the
interlayer field $E^q_3$ is essentially fixed at $E^q_3 = 0$. Nonzero    
values of $E^q_3$ cost energy of order $g_3$. As a result, the
interlayer tunneling terms $H_{tz(z+1)}$ and interlayer flux terms
$\cos(\Delta_i A^{q}_{xz,3}-A^{q}_{x z,i} + A^{q}_{x (z+1),i})$ in (\ref{multlayer}) 
are suppressed, and can be dropped from the Hamiltonian. At low energies, the
physics is then described by 
$H = \sum_z (H_{zA} + H_{zt})$ - the effective theory for \emph{decoupled}
$\nu = 1/3$ states. 

On an intuitive level, the basic physics is that large interlayer gauge fluctuations 
prohibit partons from tunneling between the layers, leading to a 
decoupled layer state. One can also think about the transition between the 3D state
and the decoupled  layer state in terms of Higgs condensation. 
Consider, for example, the bilayer case. Recall that the interlayer gauge 
field terms (\ref{bilaygauge}) in the bilayer parton Hamiltonian are given 
by
\begin{eqnarray}
H_{12A} &=& \sum_{x q} \frac{g_3}{2} (E^{q}_{x1,3})^2 \nonumber \\
&-& \sum_{xiq} J_3 \cos(\Delta_i A^{q}_{x1,3} - A^{q}_{x 1, i} + A^{q}_{x 2, i})
\end{eqnarray}
Let us view the operator $e^{iA^q_{x1,3}}$ as the creation 
operator of a boson at site $x$, while $E^q_{x1,3}$ is the number 
operator which measures the number of bosons at site $x$. The first term 
is then a potential energy term which describes the energy associated with 
having a certain number of bosons on a given site, while the second term 
is a kinetic energy term which describes a boson hopping from one site to 
a neighboring site. The presence of the combination $A^q_{1,i} - 
A^q_{2,i}$ in the argument of the cosine tells us that the boson is 
minimally coupled to the gauge field $A^q_{1,i} - A^q_{2,i}$.

It is illuminating to think about the strong and weak gauge fluctuation 
regimes in this language. When $g_3 \gg J_3$, the boson is massive (e.g. 
in a Mott insulating phase) and is therefore irrelevant at low energies. 
The interlayer gauge terms can then be dropped and the low energy effective 
Hamiltonian (\ref{bilayer}) consists of two decoupled layers, $H 
=\sum_{z=1}^2 (H_{zt} + H_{zA})$. The ground state is thus two decoupled 
$\nu = 1/3$ states.

On the other hand, when $g_3 \ll J_3$, the boson condenses. Since the boson 
is minimally coupled to the gauge field $A^q_{1,i} - A^q_{2,i}$, this 
boson condensation is a kind of Higgs condensation where the Higgs boson is 
coupled to a Chern-Simons gauge field. \cite{WW9301} When such a Higgs boson condenses, 
the result is another gapped FQH state - in this case, the bilayer parton state. 

Because one can tune from the decoupled state to the parton state by 
changing a single coupling constant $g_3$, one can speculate that these 
states are in some sense neighboring or proximate phases. This is one of 
the reasons that we propose the parton state as a candidate for an 
intermediate tunneling phase.

\section{Edge (and surface) states}
\label{edgesect}
In this section, we discuss the edge states for the $N$-layer system, both 
for finite $N$ and in the 3D limit $N \rightarrow \infty$.

First, consider the case of finite $N$. In this case, the edge theory can 
be read off from the $K$-matrix and charge vector $t$ described in section 
\ref{Nlayersect}, using the standard formalism. \cite{W9505} The result is a 
chiral boson theory 
with $3N-2$ modes, $\phi^I, I = 1, ... ,3N-2$. The Lagrangian is of the form
\begin{equation}
L = \frac{1}{4\pi} ( K_{IJ} \partial_t \phi^I \partial_x \phi^J - V_{IJ} 
\partial_x \phi^I \partial_x \phi^J)
\end{equation}
where $V_{IJ}$ is a positive definite velocity matrix which describes the 
velocities of each of the modes and the density-density interactions 
between different modes. Quasiparticle excitations are parameterized by 
integer vectors $l$ and are created by operators of the form $\exp(i l_I 
\phi^I)$. The electric charge corresponding to a quasiparticle $l$ is given 
by $q = e \cdot (t^T K^{-1} l)$.

As many aspects of the edge theory depend on microscopic details of the 
edge and can be affected by edge reconstruction, let us discuss two simple 
quantities which are universal. The first quantity - the electric Hall 
conductance - is given by
\begin{equation}
\sigma_{xy} = (t^T K^{-1} t) \cdot \frac{e^2}{h} = \frac{N e^2}{3 h}
\end{equation}
Of course, this is exactly what we expect since the parton state has $\nu = 1/3$ per 
layer.

A more interesting quantity is the thermal Hall conductance.
Recall that each chiral boson edge mode gives a contribution
of $\pm \frac{\pi^2 k_B^2}{3h} T$ to the thermal Hall conductance, with  
the sign determined by the chirality of the mode. Thus, the thermal Hall 
conductance can be computed by counting the number of edge modes. Since  
the $K$-matrix has $3N-2$ positive eigenvalues and no negative eigenvalues,
there are $3N-2$ modes propagating in one direction and no modes   
propagating in the opposite direction. We conclude that the thermal Hall
conductance is $3N-2$ (in units of $\frac{\pi^2 k_B^2}{3h} T$).      

Before concluding this section, let us briefly discuss the 3D limit, $N \rightarrow \infty$. In this case, 
the boundary is two-dimensional, and the edge states are actually 
\emph{surface} states. Let us focus on the most interesting kind of 
boundary: a boundary in the $xz$ plane (with layers oriented in the 
$xy$ plane).

The analysis of the surface states is complicated by the fact that the 
bulk has gapless modes. Because of this, we will not analyze the surface 
in detail, but rather sketch the basic qualitative picture which 
is evident in mean field theory. In mean field theory, the surface modes 
are given by $3$ species of non-interacting 2D fermions, which are chiral 
in the $x$ direction but non-chiral in the $z$ direction. The modes form 
a ``sheath" of chiral fermions with Fermi surface $k_x \sim t_z \cos(k_z a)$. 
\cite{BF9682} When one goes beyond mean field 
theory and includes gauge fluctuations, these fermions will become 
minimally coupled to the bulk ``photon" mode. The gauge fluctuations will 
certainly affect the surface theory; however, we know that they cannot gap 
out the surface modes entirely. Indeed, the gaplessness of the edge modes 
is protected by the nonzero electric Hall conductivity in the bulk
($e^2/3h$ per layer). The surface sheath therefore resembles a 
chiral analog of the Halperin-Lee-Read state \cite{HLR9312} which is protected 
against gap-forming instabilities by the topological character of the bulk phase. 

These surface states may provide the simplest experimental signature of the
$3D$ parton states. In particular, consider the $z$-axis surface longitudinal 
conductance $\sigma_{zz}$. In mean field theory, $\sigma_{zz}$ behaves just like the 
conductance of the layered integer quantum Hall system studied in \cite{BF9682}. Thus, 
$\sigma_{zz} \sim \text{const.}$ as $T \rightarrow 0$. Including gauge fluctuations,
we expect that this constant will be renormalized, but $\sigma_{zz}$ will remain
finite at zero temperature. This should be contrasted with the behavior of $\sigma_{zz}$ 
in the decoupled $\nu = 1/3$ Laughlin state. In that case $\sigma_{zz} \sim T^3$ as 
$T \rightarrow 0$. \cite{BF9682} Thus, a measurement of $\sigma_{zz}$ could, in 
principle, distinguish the $3D$ parton state from the decoupled Laughlin state.

\section{Conclusion}
In this paper, we have constructed a candidate state for a multilayer FQH 
system with average filling $\nu = 1/3$ per layer. We have proposed that 
the state may be realized in the ``intermediate tunneling regime" where the 
interlayer tunneling strength is of the same order as the Coulomb energy 
$e^2/l_B$. Our construction is based on a slave particle approach, known 
as the ``parton construction."

We have analyzed the state for both a finite number of layers $N$, and in 
the 3D limit, $N \rightarrow \infty$. In the case of a finite number of 
layers $N$, the state is a gapped FQH state and is described 
by a $(3N-2) \times (3N-2)$ $K$-matrix. Its quasiparticle excitations are 
anyonic and carry charge $e/3$.

In the $3D$ limit, $N \rightarrow \infty$, the state is more unusual. It 
supports two types of excitations: gapped $e/3$ fermionic quasiparticle 
excitations and gapless neutral collective modes. The quasiparticle 
excitations are truly 3D quasiparticles and can propagate freely both 
within and between layers. (This is in contrast to the charge $e/3$ 
excitations in the multilayer Laughlin state, which are confined to 
individual layers). The gapless neutral collective modes are 
emergent ``photon" modes originating from the slave particle gauge theory.
Unlike Maxwell photons, they come in only one polarization, and have an 
anisotropic dispersion $\omega_{\bf k}^2 \sim {\bf k}_\perp^2 + k_z^4$. 
The $e/3$ fermionic quasiparticles are minimally coupled to these ``photon" 
modes so that they have long range interactions.

The edge physics of the finite layer and 3D systems is also interesting.
When $N$ is finite, the edge theory is described by a conformal 
field theory with $3N-2$ chiral boson modes. In the 3D limit, the edge
modes are more complex. In mean field theory, the edge (or more accurately, 
surface) modes are described by three different species of non-interacting 
2D fermions which propagate chirally in the $x$-direction (e.g. the 
direction parallel to the layers) and non-chirally in the $z$ direction 
(e.g. the direction perpendicular to the layers). Going beyond mean field 
theory, we expect that these fermions are minimally coupled to the gapless 
bulk ``photon" modes. However, we have not analyzed the surface physics
in detail. This is an interesting direction for future research.

Another direction for future research would be to construct other types
of layered FQH states. For example, it would be interesting to build 
multilayer states with average filling $\nu = 1/2$ per layer - in 
particular, multilayer states which are related to the Moore-Read $\nu = 
1/2$ state \cite{MR9162} or the composite Fermi liquid $\nu = 1/2$ state 
\cite{HLR9312} (instead of the 
Laughlin $\nu = 1/3$ state which we have investigated here). One possible 
approach for this problem would be to employ a parton construction where 
one writes an electron as $c = d_1 d_2 f$, where $d_1, d_2$ are fermionic 
partons carrying charge $e/2$ and $f$ is a neutral fermionic parton. One 
could then consider mean field states where, in each layer, $d_1, d_2$ are 
in integer quantum Hall states and $f$ is in a $p+ip$ superconducting 
state or a Fermi liquid state. In this way, one may be able to construct 
multilayer and 3D relatives of the Moore-Read or composite Fermi 
liquid states. 

In general, there are clearly many possibilities for 3D multilayer FQH 
states, most of which have not been explored. We hope that the parton 
construction provides a useful tool for constructing and analyzing these 
new states of matter.

\acknowledgments
This research was supported by the Microsoft Corporation and by NSF grant 
DMR-05-29399.
\bibliography{layfqh}
\end{document}